\documentstyle[12pt]{article}
\begin{document}
\topmargin 0pt
\oddsidemargin -3.5mm
\headheight 0pt
\topskip 0mm
\addtolength{\baselineskip}{0.20\baselineskip}
\hfill hep-th/9702134

\hfill SOGANG-HEP 212/97


\vspace{0.5cm}

\begin{center}
{ \large \bf  Non-Abelian Proca model
            based on the improved BFT formalism}
\end{center}

\vspace{0.5cm}

\begin{center}
Mu-In Park$^{a}$ and Young-Jai Park$^{b}$,  \\
\end{center}
\begin{center}
{ Department of Physics and Basic Science Research Institute \\
Sogang University, C.P.O. Box 1142, Seoul 100-611, Korea} \\             
\end{center}

\vspace{1cm}

\begin{center}
{\bf ABSTRACT}
\end{center}

We present the newly improved Batalin-Fradkin-Tyutin (BFT) Hamiltonian formalism 
and the generalization to the Lagrangian formulation, which provide
the much more simple and transparent insight to the usual BFT method, with
application to the non-Abelian Proca model which has been an difficult problem
in the usual BFT method.
The infinite terms of the effectively first class constraints can be made to
be the regular power series forms by ingenious choice of
$X_{\alpha \beta}$ and $\omega^{\alpha \beta}$-matrices.
In this new method, the first class Hamiltonian,
which also needs infinite correction terms is obtained simply by
replacing the original variables in the original Hamiltonian with the BFT
physical variables.
Remarkably all the infinite correction terms can be expressed
in the compact exponential form. We also show that in our model
the Poisson brackets of the BFT physical variables in the extended
phase space are the same structure as the Dirac brackets of the original 
phase space variables. With the help of both our newly developed Lagrangian 
formulation and Hamilton's equations of motion, we obtain the desired
classical Lagrangian corresponding to the first class Hamiltonian which can be 
reduced to the generalized St\"uckelberg Lagrangian which is non-trivial
conjecture in our infinitely many terms involved in Hamiltonian and
Lagrangian.
\vspace{ 2cm}
\hrule
\vspace{1cm}
$^{a}$ Electronic address: mipark@physics.sogang.ac.kr

$^{b}$ Electronic address: yjpark@ccs.sogang.ac.kr
\newpage

\begin{center}
{\large \bf I. Introduction}
\end{center}

The Dirac method has been widely used in the Hamiltonian formalism [1] to quantize the first and the
second class constraint systems generally, which do and do not form a closed
constraint algebra in Poisson brackets, respectively.
However, since the resulting Dirac brackets are generally field-dependent
and nonlocal, and have a serious ordering problem,
the quantization is under unfavorable circumstances
because of, essentially, the difficulty
in finding the canonically conjugate pairs.
On the other hand, the quantization of first class  constraint systems
established by Batalin, Fradkin, and Vilkovisky (BFV) [2, 3],
which does not have the previously noted problems of the Dirac method
from the start,
has been well appreciated in a gauge invariant manner with preserving
Becci-Rouet-Stora-Tyutin (BRST) symmetry [4, 5].
After their works,
this procedure has been generalized to include the second class constraints
by Batalin, Fradkin, and Tyutin (BFT) [6, 7]
in the canonical formalism, and applied to various models
$^{8-10}$ obtaining the Wess-Zumino (WZ) actions [11, 12].

Recently, the BFT Hamiltonian method [7] has been systematically applied to
several linear [13] or non-linear [14] second class constraint systems
producing the
interesting result to find the new type of the WZ action which cannot be
obtained in the usual path-integral framework. It is interesting to note that
even the non-linear system requiring infinite iterations can be exactly solved
in this method by some ingenious choice of the arbitrariness in the definition of 
($ X_{\alpha \beta},\omega^{\alpha \beta} $). All these works are based on the systematic
construction of the first class Hamiltonian as a solution of the strongly
involutive relation with the effectively first constraints. However, this method
may not be easy one to much more complicated systems due to the complexity of
the routine procedure [15].

In this respect we have recently suggested the improved
method finding the first class Hamiltonian much simply and more transparently
than the usual method and applied to several simple models, i.e., Abelian
Proca model [16] and Abelian chiral Schwinger model [17]. However, these models
do not show dramatically the power of our improved method due to the
simplicity of the system. In this sense it would be worth considering the
non-Abelian Proca model [18] which has not been completely solved in the usual
method [8, 15]. This is one issue which will be attacked in this paper. 
However, even in this improved method the corresponding Lagrangian formulation 
was unclear due to the appearances of the time derivatives in the Lagrangian 
[16, 17]. This is another issue on the formalism which will be also tackled in this paper.

In the present paper, we present the newly improved BFT Hamiltonian formalism
and also the generalization to the Lagrangian formulation, which provide the 
much more simple and transparent insight to the usual BFT 
method, with
application to the non-Abelian Proca model. The model is most non-trivial
model due to the necessity of infinite correction terms involving newly
introduced auxiliary fields to perform the BFT's conversion of the weakly second class
constraints into the strongly first class constraints. However this infinite
terms can be made to be regular power series forms by ingenious choice of the $X_{\alpha \beta}$
and $\omega^{\alpha \beta}$-matrices such that our formulation can be also
applied in this case.

In Sec. II, we apply the usual BFT formalism [7] to the non-Abelian
Proca model in order to convert the weakly second class constraint system
into a strongly first class one by introducing new auxiliary fields. We find
that the effectively first class constraints needs the infinite terms involving
the auxiliary fields which being regular power series forms in our 
ingenious choice.

In Sec. III, according to our new method, the first class Hamiltonian, which also
needs infinite correction terms, is obtained simply by replacing the original
variables in the original Hamiltonian with the BFT physical variables defined 
in the extended phase space. The
effectively first class constraints can be also understood similarly. It is
proved that the infinite terms in the BFT physical variables and hence first
class constraints and the first class Hamiltonian can be expressed by
the compact exponential form. We also show that in our
model the Poisson brackets of the BFT physical variables fields in the
extended phase space are the same as the Dirac brackets of the phase space
variables in the original second class constraint system  only with replacing
the original phase variables by the BFT physical variables.

In Sec. IV, we develop the Lagrangian formulation to obtain the first
class constraints which complement and provide much more transparent insight
to the Hamiltonian formulation. Based on this Lagrangian formulation we
directly obtain the St\"uckelberg Lagrangian as the classical Lagrangian 
corresponding to the first class Hamiltonian which is most non-trivial 
conjecture related to Dirac's conjecture [1] in our infinitely many terms
involved model. This is also confirmed by
considering the Hamilton's equations of motion with some modification term in 
Hamiltonian which is proportional to  the first class constraints $\tilde{\Theta}_2^a$.
This disproves the recent argument on the in-equivalence of the St\"uckelberg and BFT formalism.

Sec. V is devoted to our conclusion and several comments

\begin{center}
{\large \bf II. Conversion from second to first class constraints }
\end{center}

Now, we first apply the usual BFT formalism which assumes the Abelian
conversion of the second class constraint of the original system [7] to
the non-Abelian Proca model of the massive photon in four dimensions
[8, 15, 18, 19], whose dynamics are described by
\begin{equation}
S = \int d^4x~
             [ -\frac{1}{4} F^{a}_{\mu\nu} F^{a, \mu\nu}
             + \frac{1}{2} m^2 A_\mu^{a} A^{a, \mu} ],
\end{equation}
where $F_{\mu\nu}^{a} = \partial_\mu A_\nu^{a} - \partial_\nu A_\mu^{a}+g f^{abc}A_{\mu}^{b} A_{\nu}^{c}$, and
$g_{\mu\nu} = diag(+,-,-,-)$.

The canonical momenta of gauge fields are given by
\begin{eqnarray}
\pi_{0}^{a} & \equiv & \frac{\delta S}{\delta \dot{A}^{a,0}} \approx 0, \nonumber\\
\pi_{i}^{a} & \equiv & \frac{\delta S}{\delta \dot{A}^{a,i}} = F_{i0}^{a}
\end{eqnarray}
with the Poisson algebra $\{ A^{a, \mu}(x), \pi^{b}_{\nu }(y) \}
=\delta ^{\mu}_{\nu} \delta_{ab} \delta^3 ({\bf x}-{\bf y})$.
The weak equality ` $\approx$ ' means that the equality is not applied before
all involved calculations are finished. [1] In contrast, the strong
equality ` $=$ ' means that the equality can be applied at all the steps of the
calculations.

Then, $\Theta_1^{a} \equiv \pi_0^{a} \approx 0$ is a primary constraint [1].
On the other hand, the total Hamiltonian is
\begin{equation}
H_{T}=H_{c}+\int d^{3}x u^{a} \Theta_{1}^{a}
\end{equation}
with the Lagrangian multipliers $u^{a}$ and the canonical Hamiltonian
\begin{equation}
H_c = \int d^3x \left[
                \frac{1}{2} (\pi_i^{a})^2 + \frac{1}{4} F_{ij}^{a} F^{a, ij}
              + \frac{1}{2} m^2 \{ (A^{a,0})^2 + (A^{a,i})^2 \}
              - A_0^{a} \Theta_2^{a}
              \right],
\end{equation}
where $\Theta_2^{a}$ is the Gauss' law constraints,
which come from the time evolution of $\Theta_1^{a}$ with $H_{T}$, defined by
\begin{eqnarray}
\Theta_2^{a} &\equiv& \partial^i \pi_i^{a}+ g f^{abc} A^{b,i} \pi_{i}^{c} + m^2 A^{a,0}  \nonumber \\
             &=&(D^i \pi_i)^a + m^2 A^{a,0} \approx 0.
\end{eqnarray}
Note that the time evolution of the Gauss' law constraints with $H_{T}$
generates no more
additional constraints if we choose only determines the multipliers as
$ u^{a} \approx -\partial_{i}A^{a,i}$. As a result, the full constraints of 
this model are
$\Theta_{\alpha}~(\alpha = 1, 2)$ which satisfy the second class constraint algebra as follows
\begin{eqnarray}
\Delta_{\alpha \beta}^{ab}(x,y) \equiv \{ \Theta_{\alpha}^{a}(x), \Theta_{\beta}^{b}(y) \}
               = \left(\begin{array}{cc}
                 0 & -m^{2} \delta^{ab}    \\
                 m^{2}\delta^{ab} & g f^{abc} \Omega^{c}
                 \end{array}
                 \right)
                \delta^3({\bf x}-{\bf y}),
\end{eqnarray}
where we denote $x=(t,\bf{x})$ and three-space vector ${\bf x}=(x^1,x^2,x^3)$
and $\epsilon_{12}=-\epsilon_{21}=1$, and $\Omega^{c}=(D^{i} \pi_{i})^{c}$.
Here we note that it is dangerous to use the constraints $\Theta_2^a \approx 0$
of Eq. (5) at this stage like as in Ref. [15] since in our case
${\tilde \Theta}^a_{\alpha} \approx 0$ of Eq. (9) only can be imposed.

Now, we introduce new auxiliary fields $\Phi^{\alpha}_a$ to convert the second
class constraints $\Theta^a_{\alpha}$ into first class ones
in the extended phase space with the fundamental Poisson algebra
\begin{eqnarray}
   \{ {\cal F}, \Phi^{b,i} \} &=& 0,  \nonumber \\
   \{ \Phi^{\alpha}_{a}(x), \Phi^{\beta}_{b}(y) \} = \omega^{\alpha \beta}_{ab}(x,y) &=&
                      -\omega^{\beta \alpha}_{ba}(y,x),
\end{eqnarray}
where ${\cal F}=(A^{a}_{ \mu}, \pi^{a}_{ \mu})$.
Here, the constancy, i.e., the field independence of $\omega^{\alpha \beta}_{ab}(x,y)$, is
considered for simplicity.

According to the usual BFT method [7],
the modified constraints $\widetilde{\Theta}_{\alpha}^{a}$ with the property
\begin{eqnarray}
\{\widetilde{\Theta}_{\alpha}^a, \widetilde{\Theta}_{\beta}^b \}=0,
\end{eqnarray}
which is called the ${\it Abelian~ conversion}$, which is rank-0, of the second
class constraint (6) are generally given by
\begin{equation}
  \widetilde{\Theta}_{\alpha}^a[{\cal F} ; \Phi]
         =  \Theta_{\alpha}^a + \sum_{n=1}^{\infty} \widetilde{\Theta}_{\alpha}^{a (n)}=0;
                       ~~~~~~\tilde{\Theta}_{\alpha}^{(n)} \sim (\Phi)^n
\end{equation}
satisfying the boundary conditions,
$\widetilde{\Theta}_{\alpha}^a[ {\cal F};  0] = \Theta_{\alpha}^a$.
Note that the modified constraints $\widetilde{\Theta}_{\alpha}^a$
become strongly zero by introducing the auxiliary fields $\Phi_{\alpha}^a$,
i.e., enlarging the phase space,
while the original constraints $\Theta^a_{\alpha}$ are weakly zero.
As will be shown later, essentially due to this property, the result of the
Dirac formalism can be easily read off from the BFT formalism.
The first order correction terms in the infinite series [7] are
simply given by
\begin{equation}
  \widetilde{\Theta}_{\alpha}^{a (1)}(x) = \int d^3 y X_{\alpha \beta}^{ab} (x,y)\Phi^{\beta}_{b} (y),
\end{equation}
and the first class  constraint algebra (8) of $\widetilde{\Theta}_{\alpha}^a$ requires
the following relation
\begin{equation}
   \triangle_{\alpha \beta}^{ab} (x,y) +
   \int d^3 w~ d^3 z~
        X_{\alpha \gamma}^{ac} (x,w) \omega^{\gamma \delta}_{cd} (w,z) X_{\beta \delta}^{bd} (y,z)
         = 0.
\end{equation}
However, as was emphasized in Refs. [13 - 15],
there is a natural arbitrariness
in choosing the matrices $\omega^{\alpha \beta}_{ab}$ and $X_{\alpha \beta}^{ab}$ from Eqs. (7) and (10),
which corresponds to canonical transformation
in the extended phase space [6, 7].
Here we note that Eq. (11) can not be
considered as the matrix multiplication exactly unless $X_{\beta \delta}^{bd} (y,z)$ has
some symmetry, $X_{\beta \delta}^{bd} (y,z)=k X_{\alpha \beta}^{db} (z,y)$ with constant $k$ because
of the form of the last two
product of the matrices $\int d^3z \omega^{\gamma \delta}_{cd} (\omega, z) X_{\beta \delta}^{bd} (y,z)$ in
the right hand side of Eq. (11). Thus, using this arbitrariness
we can take the
simple solutions without any loss of generality, which are compatible with
Eqs. (7) and (11) as
\begin{eqnarray}
  \omega^{\alpha \beta}_{ab} (x,y)
         &=&  \epsilon^{\alpha \beta}\delta^{ab} \delta^3({\bf x}-{\bf y}), \nonumber  \\
  X_{\alpha \beta}^{ab} (x,y)
         &=&  m
         \left( \begin{array}{cc}
         0 & -\delta^{ab} \\
         \delta^{ab} & (g/2 m^2) f^{abc} \Omega^c
         \end{array}
         \right)
          \delta^3({\bf x}-{\bf y}),
\end{eqnarray}
i.e., antisymmetric $\omega^{ij}(x,y)$ and $X_{ij}(x,y)$ such that
Eq. (11) is the form of the matrix multiplication exactly [10, 13, 15]
\begin{eqnarray*}
\Delta_{\alpha \beta}^{ab} (x,y) - \int d^3 \omega d^3 z X_{\alpha \gamma}^{ac} (x, \omega) \omega^{\gamma \delta}_{cd}
(\omega, z) X_{\delta \beta}^{db} (z,y) =0. \nonumber
\end{eqnarray*}
Note that $X_{\alpha \beta}^{ab} (x,y)$ needs not be generally antisymmetric, while
$\omega^{\alpha \beta}_{ab} (x,y)$ is always antisymmetric by definition of Eq. (7).
However, the symmetricity or antisymmetricity of $X_{\alpha \beta}^{ab}(x,y)$ is, by experience, a powerful property for
the solvability of (9) with finite iteration [13] or with
infinite {\it regular } iterations [14].  Actually this solution provides the latter case in our model as will be shown later.

Now with this proper choice,
the first order corrections of the modified constraints become
\begin{eqnarray}
\widetilde{ \Theta}_{1}^{a(1)}&=&-m \Phi^{2}_{a}, \nonumber \\
\widetilde{ \Theta}_{2}^{a(1)}&=&m \Phi^{1}_{a}+\frac{g}{2m}f^{abc} \Phi^{2}_{b} \Omega^{c} \nonumber \\
                              &=&m \Phi^{1}_{a}+\frac{g}{2m}(\Phi^{2} \times \Omega)^{a},
\end{eqnarray}
where the cross operation `$\times $' represents $(A \times B)^a =f^{abc} A^b B^c$.
Note that $\widetilde{\Theta}_2^{a(1)} $ have the non-Abelian correction terms
contrast to $\widetilde{\Theta}_1 ^{a(1)}$.

The higher order iteration terms, by omitting the spatial coordinates for simplicity, [7]
\begin{eqnarray}
\widetilde{\Theta}^{a (n+1)}_{\alpha} =
                  - \frac{1}{n+2} \Phi^{\delta}_d
                  \omega_{\delta \gamma}^{dc} X^{\gamma \beta}_{cb} B_{\beta \alpha}^{ba (n)}~~~~~~~(n \geq 1)
\end{eqnarray}
with
\begin{eqnarray}
B^{ba(n)}_{\beta \alpha} \equiv \sum^{n}_{m=0}
          \{ \widetilde{\Theta}^{b(n-m)}_{\beta},
                \widetilde{\Theta}^{a(m)}_{\alpha} \}_{({\cal F})}
        + \sum^{n-2}_{m=0}
              \{ \widetilde{\Theta}^{b(n-m)}_{\beta},
                     \widetilde{\Theta}^{a(m+2)}_{\alpha} \}_{(\Phi)}
\end{eqnarray}
are found to be non-vanishing contrast to the Abelian  model [15, 16].
Here, $\omega_{\delta \gamma}^{dc}$ and $X^{\gamma \beta}_{dc}$ are the inverse of
$\omega^{\delta \gamma}_{dc}$ and $X_{\gamma \beta}^{dc}$,
and the Poisson brackets including the subscripts are defined by
\begin{eqnarray}
\{A,B \}_{(Q,P)} &\equiv& \frac{\partial A}{\partial Q}
                  \frac{\partial B}{\partial P}
                - \frac{\partial A}{\partial P}
                  \frac{\partial B}{\partial Q},  \nonumber \\
\{A,B \}_{(\Phi)} &\equiv& \sum_{\alpha ,\beta}
               \left[  \frac{\partial A}{\partial \Phi^{\alpha}}
                       \frac{\partial B}{\partial \Phi^{\beta}}
                     - \frac{\partial A}{\partial \Phi^{\beta}}
                       \frac{\partial B}{\partial \Phi^{\alpha}} \right],
\end{eqnarray}
where $(Q,P)$ and $(\Phi^{\alpha}, \Phi^{\beta} )$ are the conjugate pairs
of the original and auxiliary fields,
respectively. After some calculations it is not difficult to prove that
\begin{eqnarray}
\widetilde{\Theta}^{a (n+1)}_{1}& =& 0, \nonumber \\
\widetilde{\Theta}^{a (n+1)}_{2} &=& \frac{1}{(n+2)!}
  \left(\frac{g}{m} \right)^{n+1} f^{a b_{1} c_{1} } f^{c_{1} b_{2} c_{2} } \cdots f^{c_{n} b_{n+1} g} \Phi^{2}_{b_{1}} \Phi^{2}_{b_{2}} \cdots \Phi^{2}_{b_{n+1}} \Omega^{g} \nonumber \\
              &=& \frac{1}{(n+2)!} \left(\frac{g}{m} \right)^{n+1} [\Phi^{2} \times ( \Phi^{2} \times ( \cdots ( \Phi^{2} \times \Omega )\cdots)) ]_{\mbox{(n+1)-fold}}^{a} ~~~~(n \geq 1)
\end{eqnarray}
using the mathematical induction. Here `$n$-fold' means that the number of
the parenthesis and the bracket is $n$. Note that all the higher order corrections
$\widetilde{\Theta}_{2}^{a(n+1)}$ are essentially the non-Abelian effect.

Hence in our model with the proper choice of Eq. (12) the effectively first class constraints to all orders become
\begin{eqnarray}
\widetilde{\Theta}^{a}_{1}& =& \Theta_{1}^a- m \Phi^{2}_{a}, \nonumber \\
\widetilde{\Theta}^{a}_{2} &=& \Theta_{2}^{a} +m \Phi^{1}_{a}+ \sum_{n=1}^{\infty}
               \frac{1}{(n+1)!} \left(\frac{g}{m} \right)^{n} [\Phi^{2} \times ( \Phi^{2} \times ( \cdots ( \Phi^{2} \times \Omega )\cdots)) ]_{(\mbox{n-fold})}^{a}.
\end{eqnarray}
with $[\Phi^{2} \times ( \Phi^{2} \times ( \cdots ( \Phi^{2} \times \Omega )\cdots)) ]_{\mbox{0-fold}}^{a} \equiv \Omega^a$.
\vspace{1cm}

\begin{center}
{\large \bf III. Physical variables, first class Hamiltonian, and Dirac brackets }
\end{center}

Now, corresponding to the original variables ${\cal F}$,
the ${\it physical }$ variables, called {\it BFT physical variables}
in the extended phase space, within the Abelian conversion,
$\widetilde{\cal F}$, which are strongly involutive, i.e.,
\begin{eqnarray}
\{ \widetilde{\Theta}_{\alpha}^a, \widetilde{\cal F} \} =0
\end{eqnarray}
can be generally found as
\begin{eqnarray}
\widetilde{\cal F}[{\cal F}; \Phi ] =
          {\cal F} + \sum^{\infty}_{n=1} \widetilde{\cal F}^{(n)},
        ~~~~~~~ \widetilde{\cal F}^{(n)} \sim (\Phi)^{n}
\end{eqnarray}
satisfying the boundary conditions,
$\widetilde{\cal F}^{a, \mu}[{\cal F}; 0 ] = {\cal F}$.
Here, the first order iteration terms which are given by the formula
\begin{eqnarray}
\widetilde{{\cal F}}^{(1)}=
                   - \Phi^{\beta}\omega_{\beta \gamma} X^{\gamma \delta}
                   \{ \Theta_{\delta}, {\cal F} \}_{({\cal F})}  \nonumber  \\
\end{eqnarray}
become as follows
\begin{eqnarray}
\widetilde{A}^{a(1)}_{0}&=&\frac{1}{m} \Phi^{1}_{a} -\frac{g}{2 m^3} (\Phi^2 \times \Omega)^a, \nonumber \\
\widetilde{A}^{a(1)}_{i}&=&\frac{1}{m} [-\partial_{i}\Phi^{2}_{a} +g (\Phi^2 \times A_{i})^a], \nonumber \\
\widetilde{\pi}^{a(1)}_{i}&=&\frac{g}{m}  (\Phi^2 \times \pi_{i})^a, \nonumber \\
\widetilde{\pi}^{a(1)}_{0}&=&\Theta^{a(1)}_1.
\end{eqnarray}

The remaining higher order iteration terms which are given by general formula [7]
\begin{eqnarray}
\widetilde{{\cal F}}^{(n+1)} = -\frac{1}{n+1}
   \Phi^{\beta}\omega_{\beta \gamma} X^{\gamma \delta} (G_F)^{(n)}_{\delta}
\end{eqnarray}
with
\begin{eqnarray}
(G_F)^{(n)}_{\delta} = \sum^{n}_{m=0}
           \{ \Theta_{\delta}^{(n-m)}, \widetilde{{\cal F}}^{ (m)}\}_{({\cal F})}
         +   \sum^{n-2}_{m=0}
            \{ \Theta_{\delta}^{(n-m)}, \widetilde{{\cal F}}^{(m+2)}\}_{(\Phi)}
           +  \{ \Theta_{\delta}^{(n+1)}, \widetilde{{\cal F}}^{(1)} \}_{(\Phi)},
\end{eqnarray}
are also found to be as follows after some routine calculations
\begin{eqnarray}
\widetilde{A}^{a (n+1)}_{0} &=& -\frac{n+1}{(n+2)!} \frac{1}{m^2}
  \left(\frac{g}{m} \right)^{n+1} [\Phi^{2} \times ( \Phi^{2} \times ( \cdots ( \Phi^{2} \times \Omega )\cdots)) ]_{\mbox{(n+1)-fold}}^{a},  \nonumber \\
\widetilde{A}^{a (n+1)}_{i} &=& \frac{1}{(n+1)!}
  \left(\frac{g}{m} \right)^{n+1} [\Phi^{2} \times ( \Phi^{2} \times ( \cdots ( \Phi^{2} \times A_{i} )\cdots)) ]_{\mbox{(n+1)-fold}}^{a} \nonumber \\
&&-\frac{1}{(n+1)!} \frac{1}{g}
  \left(\frac{g}{m} \right)^{n+1} [\Phi^{2} \times ( \Phi^{2} \times ( \cdots ( \Phi^{2} \times \partial_{i} \Phi^2  )\cdots)) ]_{\mbox{n-fold}}^{a}, \nonumber \\
\widetilde{\pi}^{a (n+1)}_{i} &=& \frac{1}{(n+1)!}
  \left(\frac{g}{m} \right)^{n+1} [\Phi^{2} \times ( \Phi^{2} \times ( \cdots ( \Phi^{2} \times \pi_{i} )\cdots)) ]_{\mbox{(n+1)-fold}}^{a}, \nonumber \\
\widetilde{\pi}^{a (n+1)}_{0} &=& \widetilde{\Theta}_{1}^{a(n+1)} =0.
\end{eqnarray}

Hence, the BFT physical variables in the extended phase space
are finally found to be
\begin{eqnarray}
\widetilde{A}^{a }_{0} &=& A_{0}^a +\frac{1}{m} \Phi^{1}_a -\sum_{n=1}^{\infty} \frac{n}{(n+1)!} \frac{1}{m^2}
  \left(\frac{g}{m} \right)^{n} [\Phi^{2} \times ( \Phi^{2} \times ( \cdots ( \Phi^{2} \times \Omega )\cdots)) ]_{\mbox{n-fold}}^{a},  \nonumber \\
\widetilde{A}^{a}_{i} &=& A_{i}^a +\sum_{n=1}^{\infty} \frac{1}{n!}
  \left(\frac{g}{m} \right)^{n} [\Phi^{2} \times ( \Phi^{2} \times ( \cdots ( \Phi^{2} \times A_{i}  )\cdots)) ]_{\mbox{n-fold}}^{a} \nonumber \\
   &&-\sum_{n=1}^{\infty} \frac{1}{n!} \frac{1}{g}
  \left(\frac{g}{m} \right)^{n} [\Phi^{2} \times ( \Phi^{2} \times ( \cdots ( \Phi^{2} \times \partial_{i} \Phi^2 )\cdots)) ]_{\mbox{(n-1)-fold}}^{a}, \nonumber \\
  &=& -\sum_{n=1}^{\infty} \frac{1}{n!}
  \left(\frac{g}{m} \right)^{n} [\Phi^{2} \times ( \Phi^{2} \times ( \cdots ( \Phi^{2} \times \partial_{i} \Phi^2 )\cdots)) ]_{\mbox{(n-1)-fold}}^{a} \nonumber \\
      &&+ \left[exp \{ \frac{g}{m} \Phi^2 \times \} A_{i} \right]^a, \nonumber \\
\widetilde{\pi}^{a}_{i} &=& {\pi}^{a}_{i} + \sum_{n=1}^{\infty} \frac{1}{n!}
  \left(\frac{g}{m} \right)^{n} [\Phi^{2} \times ( \Phi^{2} \times ( \cdots ( \Phi^{2} \times \pi_{i} )\cdots)) ]_{\mbox{n-fold}}^{a} \nonumber \\
   &=& \left[exp \{ \frac{g}{m} \Phi^2 \times \} \pi_{i} \right]^a, \nonumber \\
\widetilde{\pi}^{a }_{0} &=& \widetilde{\Theta}_{1}^{a},
\end{eqnarray}
by defining the exponential of the $\times$-operation as follows
\begin{eqnarray}
exp\{A \times \} B&=&B +A \times B +\frac{1}{2!}(A \times ( A \times B)) + \cdots \nonumber \\
               &=&B +\sum_{n=1}^{\infty} \frac{1}{n!}[A \times ( A \times ( \cdots (A \times B)\cdots))]_{\mbox{n-fold}}.
\end{eqnarray}
Here, we note that the appearance of the infinite terms of correction are
independent on the (non-Abelian) gauge group such that a genuine property of
the Abelian conversion of the non-Abelian model of Proca theory. However in
this form the gauge transformation property of the BFT physical variables is
not so transparent. Hence we needs to consider more compact expressions to
achieve this goal. To this end, we, first of all, note that the following
relation
\begin{eqnarray}
&&[i \phi, [i \phi, [ \cdots [ i \phi, M] \cdots]]]_{\mbox{n-fold}} \nonumber \\
&& =(-1)^{n} [ \phi \times ( \phi \times ( \cdots ( \phi \times M)\cdots))]_{\mbox{n-fold}}
\end{eqnarray}
is satisfied such that the following useful formula are produced as
\begin{eqnarray}
e^{i \phi} M e^{-i \phi} &=&
\sum_{n=0}^{\infty} \frac{1}{n !} [i \phi, [i \phi, [ \cdots [ i \phi, M] \cdots]]]_{\mbox{n-fold}} \nonumber \\
&=& \sum_{n=0}^{\infty} \frac{(-1)^{n}}{n !} [ \phi \times ( \phi \times ( \cdots ( \phi \times M)\cdots))]_{\mbox{n-fold}} \nonumber \\
&=& exp\{- \phi \times \} M ,\\
e^{i \phi} \frac{ \partial}{\partial \alpha} e^{-i \phi} &=&
-\sum_{n=0}^{\infty} \frac{1}{(n+1) !} [i \phi, [i \phi, [ \cdots [ i \phi, i \frac{\partial}{\partial \alpha} \phi] \cdots]]]_{\mbox{n-fold}} \nonumber \\
&=& -\sum_{n=0}^{\infty} \frac{(-1)^{n}}{(n+1)!} [ \phi \times ( \phi \times ( \cdots ( \phi \times i \frac{\partial}{\partial \alpha} \phi)\cdots))]_{\mbox{n-fold}},
\end{eqnarray}
where $\phi= \phi^a T^a, M=M^a T^a, [T^a, T^b]=ig f^{abc} T^c$ for any
non-Abelian gauge group ${\cal G}$ with generator $T^a$. Then, by using these formulas (29) and (30)
$\widetilde{A}_{i}^a$ and $\widetilde{\pi}_{i}^a$ can be re-expressed more
compactly as follows
\begin{eqnarray}
\widetilde{A}_{i}^a[{\cal F}; \Phi] &=& -\frac{i}{g}\left( \partial_i W \cdot W^{-1} \right)^a + \left( W A_{i} W^{-1} \right)^a, \\
\widetilde{\pi}_{i}^a[{\cal F}; \Phi] &=&  \left( W \pi_{i} W^{-1} \right)^a
\end{eqnarray}
with the matrix valued factor $W=exp\{-\frac{ig}{m} \Phi^2 \}$.
Furthermore, it is not difficult to show the following formula, by using the form (31)
\begin{eqnarray}
\widetilde{F}_{ij}^a[{\cal F}; \Phi]&=&  \left( W F_{ij}[{\cal F}] W^{-1} \right)^a    \nonumber \\
                                      &=& F_{ij}^a [\widetilde{{\cal F}}].
\end{eqnarray}
Actually, these compact expressions of (31)--(33) can be easily expected
ones in the BFT formalism if we remind the spirit of the formalism itself. In
this formulation, the BFT physical variables $\widetilde{{\cal F}}$ satisfy the
Eq. (19), i.e.,
\begin{eqnarray}
\{ \widetilde{\Theta}^a_{\alpha}, \widetilde{{\cal F}} \}=0.
\end{eqnarray}
But, if we remember that the first class constraints
(here $\widetilde{\Theta}^a _{\alpha}$ contrast to second class constraints
${\Theta}^a _{\alpha}$) can be generators of gauge transformations, it is clear
that $\widetilde{{\cal F}}$ should be gauge invariant one in this BFT formalism. 
Actually, under the usual non-Abelian gauge transformation
\begin{eqnarray}
&&A_{\mu}^a \rightarrow {A_{\mu}^{a }}'=-\frac{i}{g} \left(\partial_{\mu} U \cdot U^{-1}\right)^a +\left(U A_{\mu} U^{-1}\right)^a,  \nonumber \\
&&F_{\mu \nu}^a \rightarrow {F_{\mu \nu}^{a}}'=\left(U F_{\mu \nu} U^{-1}\right)^a
\end{eqnarray}
the combinations of (31)-(32) are the only possible combinations of manifest gauge invariance with 
the factor $W$ transformation
\begin{eqnarray}
W \rightarrow W'=W U^{-1}.
\end{eqnarray}
This is the story of the BFT physical variables
$\widetilde{{\cal F}}$ corresponding to the original phase space variables
${\cal F}$. On the other hand, it is interesting to note that the BFT physical
variables for the auxiliary fields, which were not existed originally, become
identically zero as it should be, i.e.,
\begin{eqnarray}
\widetilde{\Phi}^{\alpha}_a=0.
\end{eqnarray}

Similar to these BFT physical variables corresponding to the fundamental variables ${\cal F}$,
the BFT physical quantities for more complicated physically interesting 
quantities can be found
by the systematic application of the usual BFT method in principle by
considering the solutions like as Eq.(19) [7, 13-15] although
the inspection solutions may not impossible in some simple cases. However, in
many cases like as in our non-Abelian model this calculation needs some routine and tedious procedure depending on the
complexity of the quantities.

In this respect, we consider our recently proposed approach using the novel 
property [7, 16, 17, 20, 21]
\begin{eqnarray}
\widetilde{K}[{\cal F}; \Phi ]= K[\widetilde{\cal F}];
\end{eqnarray}
for the arbitrary function or functional $K$ defined on the original phase
space variables unless $K$ has the time derivatives. Then
the following relation
\begin{eqnarray}
\{ K[\widetilde{\cal F}],\widetilde{\Theta}_{\alpha} \}=0
\end{eqnarray}
is automatically satisfied for
any function $K$ not having the time derivatives
because $\widetilde{\cal F}$ and their spatial
derivatives already
commute with $\widetilde{\Theta}_{\alpha}$ at equal times by definition.
However, we note that this property is not simple when time
derivatives exits because this problem depends on the definition of time
derivative. This problem will be treated in Sec. IV. and we will show that the
equality of Eq. (38) is still satisfied even with time derivatives.

On the other hand, since the solution $K$ of Eq. (39) is
unique up to the terms proportional to the first class constraints
$\widetilde{\Theta}^a_{\alpha}$, [16, 17]
\begin{eqnarray}
u_{\alpha}^a[\widetilde{{\cal F}}] \widetilde{\Theta}_{\alpha}^a,
\end{eqnarray}
$K[\widetilde{\cal F}]$ can be identified
with $\widetilde{K}[{\cal F}; \Phi^{\alpha}]$ modulus the term (40).

Using this elegant property
we can directly obtain, after some calculation, the desired first class Hamiltonian $\widetilde{H_{c}}$
corresponding to the canonical Hamiltonian $H_{c}$ of Eq. (4) as follows
\begin{eqnarray}
\widetilde{H_{c}}[{\cal F}; \Phi^{\alpha} ]
    &=&    H_{c}[\widetilde{{\cal F}}]\nonumber \\
    &=&  \int d^3x \left[
          \frac{1}{2}({\pi}_i^a)^2 
      + \frac{1}{4} ({F_{ij}^a})^{2}
      + \frac{1}{2} m^2  ({A}^{0~a}+\frac{1}{m} \Phi^{1~a}+{\cal K}^a)^2
            \right. \nonumber \\
    && ~~~~~~ \left. 
                  + \frac{m^2}{2}\left\{A^{i~a}-\frac{i}{g} (W^{-1} \partial_i W)^a \right\}^2 
      - ({A}_0^a + \frac{1}{m} \Phi^{1~a} +{\cal K}^a) \widetilde{\Theta}_2^a
              \right],  \nonumber \\
      &=&    H_{c}[{\cal F}]+
             \int d^{3}x \left[
                -\frac{m^2}{2 g^2} \left\{(\partial_{i} W^{-1} \partial_i W )^a \right\}^2
                -\frac{ i m^2}{g}A_i^a (W^{-1} \partial_i W)^a  \right. \nonumber \\
      && ~~~~~~\left.  \frac{1}{2} (\Phi^{1~a} +m {\cal K}^a)^2
              -A_0^a(W \Omega W^{-1} -\Omega)^a
               -\frac{1}{m}(\Phi^{1~a} +m{\cal K}^a) \tilde{\Theta}^{a}_2
               \right],
\end{eqnarray}
where we have used the abbreviations
\begin{eqnarray}
{\cal K}^a &\equiv & -\sum_{n=1}^{\infty} \frac{n}{(n+1)!} \frac{1}{m^2}
  \left(\frac{g}{m} \right)^{n} [\Phi^{2} \times ( \Phi^{2} \times ( \cdots ( \Phi^{2} \times \Omega )\cdots)) ]_{\mbox{n-fold}}^a \nonumber \\
\end{eqnarray}
such that $\tilde{A}_0^a$ can be expressed as
\begin{eqnarray}
\tilde{A}_0^a =A_0^a+\frac{1}{m} \Phi^{1~a} +{\cal K}^a.
\end{eqnarray}
Then, according to Eq.(3) and the property (38), the first class
Hamiltonian for the total Hamiltonian $H_T$ becomes
\begin{eqnarray}
\widetilde{H}_{T}[A^{\mu}, \pi_{\nu}; \Phi^{i} ] =
  \widetilde{H}_{c}[A^{\mu}, \pi_{\nu}; \Phi^{i} ] -\int d^{3}x( \partial_{i} \widetilde{A}^{i})^a \widetilde{\Theta}_{1}^a.
\end{eqnarray}
Note that the difference of $\widetilde{H}_T$ and $\widetilde{H}_c$ is physically
unimportant since the difference is nothing but the ambiguity of (40) which
being inherent in the definition of the BFT physical quantities $\widetilde{K}$ [16].
On the other hand, our adopted method for obtaining $\widetilde{H}_c$ using
(38) as well as the compact forms (31)-(33) is crucial for the relatively
simple and compact result which has not been obtained so far [8, 15].

Furthermore, it is important to note that all our constraints have already this
property, i.e.,
$\widetilde{\Theta}^a_{\alpha}[{\cal F}; \Phi]
=\Theta_{\alpha}^a[\widetilde{\cal F}]$
like as follows
\begin{eqnarray}
\widetilde{\Theta}_1^a &=& \widetilde{\pi}_0^a= \Theta_1^a -m \Phi^2_a , \nonumber \\
\widetilde{\Theta}_2^{a} &=& \partial^i \widetilde{\pi}_i^{a}+ g f^{abc} \tilde{A}^{b,i} \tilde{\pi}_{i}^{c} + m^2 \tilde{A}^{a,0} , \nonumber \\
             &=& (W \Omega W^{-1})^a +m^2 (A_{0} + \frac{1}{m} \Phi^1 +{\cal K})^a.
\end{eqnarray}
In this way, the second class constraints system
$\Theta_{\alpha}^a[{\cal F}] \approx 0$ is converted into
the first class constraints one
$\widetilde{\Theta}^a_{\alpha}[{\cal F}; \Phi ] =0$
with the boundary conditions
$\widetilde{\Theta}^a_{\alpha} |_{\Phi=0}=\Theta^a_{\alpha}$.

On the other hand, in the Dirac formalism [1]
one can generally make the second class constraint system
$\Theta_{\alpha}[{\cal F}] \approx 0$ into the first class constraint one
$\Theta_{\alpha}[{\cal F}]=0$ only by deforming the phase space
$(A^{\mu}, \pi_{\mu})$ without introducing any new fields.
Hence, it seems that
these two formalisms are drastically different ones.
However, remarkably the
Dirac formalism can be easily read off from the usual BFT-formalism [7]
by noting that the Poisson bracket in the extended phase space with
$\Phi \rightarrow 0$ limit becomes
\begin{eqnarray}
\{ \widetilde{A}, \widetilde{B} \} |_{\Phi=0}
              &=&   \{A, B \}
                  - \{A, \Theta_{\alpha} \} \Delta^{\alpha \alpha '} \{\Theta_{\alpha '}, B \}  \nonumber \\
              &=&   \{A, B \}_{D}
\end{eqnarray}
where $\Delta^{\alpha \alpha '}=-X^{\beta \alpha} \omega_{\beta \beta '} X^{\beta ' \alpha '}$ is the inverse of
$\Delta_{\alpha \alpha '}$ in Eq. (6). About this remarkable relation, we note that this
is essentially due to the Abelian conversion method of the original second
class constraint. In this case the Poisson brackets between the constraints and
the other things in the extended phase space are already strongly zero
\begin{eqnarray}
\{ \widetilde{\Theta}_{\alpha}, \widetilde{A} \} &=&0 , \nonumber \\
\{ \widetilde{\Theta}_{\alpha}, \widetilde{\Theta}_{\beta} \} &=&0,
\end{eqnarray}
which resembles the property of the Dirac bracket in the non-extended phase
space
\begin{eqnarray}
\{\Theta_{\alpha}, A \}_{D} &=&0, \nonumber \\
\{\Theta_{\alpha}, \Theta_{\beta} \}_{D} &=&0,
\end{eqnarray}
such that
\begin{eqnarray}
\{ \widetilde{\Theta}_{\alpha}, \widetilde{A} \} |_{\Phi=0}
          &\equiv& \{ \Theta_{\alpha}, A \}^{*} =0, \nonumber \\
\{ \widetilde{\Theta}_{\alpha}, \widetilde{\Theta}_{\beta} \} |_{\Phi=0}
           &\equiv& \{ \Theta_{\alpha}, \Theta_{\beta} \}^{*} =0
\end{eqnarray}
are satisfied for some bracket in the non-extended phase space $\{~~,~~\}^{*}$.
However, due to the uniqueness of the Dirac bracket [22] it is natural
to expect the previous result (46) is satisfied, i.e.,
\begin{eqnarray}
\{~~,~~\}^{*} =\{~~,~~\}_{D}
\end{eqnarray}
without explicit manipulation. Moreover we add that, due to similar reason,
some non-Abelian generalization of the Abelian conversion as
\begin{eqnarray}
\{\widetilde{\Theta}_{\alpha}, \widetilde{A} \} &=&\alpha_{\alpha \beta}
\Phi^{\beta} +\alpha_{\alpha \beta \gamma } \Phi^{\beta} \Phi^{\gamma} + \cdots ,\nonumber \\
\{\widetilde{\Theta}_{\alpha}, \widetilde{\Theta}_{\beta} \} &=&\beta_{\alpha \beta \gamma } \Phi^{\gamma} +
    \beta_{\alpha \beta \gamma \delta } \Phi^{\gamma} \Phi^{\delta} + \cdots
\end{eqnarray}
also gives the same result (46) with the functions $\alpha_{\alpha \beta}, \alpha_{\alpha \beta \gamma },
 \beta_{\alpha \beta \gamma}, etc, \cdots $ of the original phase variables
 ${\cal F}$.

As an specific example,
let us consider the Poisson brackets between the phase space variables in Eq.(26).
If we calculate these brackets between the BFT physical variables, after
some manipulation we could obtain the following result
\begin{eqnarray}
&&\{\widetilde{A}^{a,0}(x), \widetilde{A}^{b,j}(y) \}
         = \frac{1}{m^{2}}(\delta^{ab} \partial_{x}^{j}-g f^{abc} \tilde{A}^{c,j} )\delta({\bf x}-{\bf y}),
                                                           \nonumber \\
&&\{\widetilde{A}^{a,0}(x), \widetilde{A}^{b,0}(y) \}=\frac{1}{m^4} g f^{abc}(\tilde{\Theta}_2^c -m^2 \tilde{A}_0^c)\delta({\bf x}-{\bf y}),
                                                            \nonumber  \\
&&\{\widetilde{A}^{a,j}(x), \widetilde{A}^{b,k}(y) \} =0,
                                                           \nonumber \\
&&\{\widetilde{\pi}_{ \mu}^a (x), \widetilde{\pi}_{\nu}^b(y) \}=0,
                                                           \nonumber \\
&&\{\widetilde{A}^{a,i}(x), \widetilde{\pi}_{j}^b(y) \}
         = \delta^{ab} \delta^{i}_{j}  \delta (\bf{x}-\bf{y}),
                                                           \nonumber \\
&&\{\widetilde{A}^{a,0}(x), \widetilde{\pi}_{j}^b(y) \}
         =\frac{1}{m^2} g f^{abc} \tilde{\pi}^c_j (x)\delta (\bf{x}-\bf{y}) ,
                                                           \nonumber \\
&&\{\widetilde{A}^{a,\mu}(x), \widetilde{\pi}_{0}^b(y) \} =0.
\end{eqnarray}
All these results can be directly calculated by using the solutions of the BFT physical
variables in Eq.(26). However, in some cases this is not easy work, but
needs routine and tedious calculation as in the determination of $\tilde{{\cal F}}$ in Eq.(26).
This is the case of the Poisson brackets involving $\tilde{A}_0^a$, i.e.,
$\{\tilde{A}_0^a, \tilde{A}^{b,j} \}, \{ \tilde{A}^{a,0}, \tilde{A}^{b,0} \}$,
and $\{\tilde{A}^{a,0}, \tilde{\pi}^b_j \}$ . In these cases the calculations
are more simplified
by expressing $\tilde{A}^{a,0}$ as $(1/m^2)[\tilde{\Theta}_2^a
-(\tilde{D}^i \tilde{\pi}_i)^a ]$ and using the Poisson brackets involving
$\tilde{A}^{a,i}$ and $\tilde{\pi}_j^b$ which can be easily calculated from the
direct use of Eq.(26).

Here, it is interesting to note firstly that all the Poisson brackets between
$\tilde{{\cal F}}$'s are expressed still by $\tilde{{\cal F}}$'s which will be
found to be important later in the
discussion of the time derivatives of the BFT physical variables. Furthermore,
the results of these Poisson brackets of $\tilde{{\cal F}}$ have exactly the same form
of the Dirac's brackets between the field ${\cal F}$ but only replacing $\tilde{{\cal F}}$
instead of ${\cal F}$ if there are field dependence in value of the Poisson brackets.
Actually these two properties are general ones from the property (46), which
can be checked explicitly in the case as follows
\begin{eqnarray}
&&\{\widetilde{A}^{a,0}(x), \widetilde{A}^{b,j}(y) \}|_{\Phi=0}
= \frac{1}{m^{2}}(\delta^{ab} \partial_{x}^{j}-g f^{abc} {A}^{c,j} )\delta({\bf x}-{\bf y})
=\{{A}^{a,0}(x), {A}^{b,j}(y) \}_{D},    \nonumber \\
&&\{\widetilde{A}^{a,0}(x), \widetilde{A}^{b,0}(y) \}|_{\Phi=0}
=\frac{1}{m^4} g f^{abc}({\Theta}_2^c -m^2 {A}_0^c)\delta({\bf x}-{\bf y})
=\{{A}^{a,0}(x), {A}^{b,0}(y) \}_{D},     \nonumber  \\
&&\{\widetilde{A}^{a,j}(x), \widetilde{A}^{b,k}(y) \}|_{\Phi=0}
=0=\{{A}^{a,j}(x), {A}^{b,k}(y) \}_{D},    \nonumber \\
&&\{\widetilde{\pi}_{\mu}^a (x), \widetilde{\pi}_{\nu}^b(y) \}|_{\Phi=0}
=0=\{{\pi}_{\mu}^{a}(x), {\pi}_{\nu}^{b}(y) \}_{D},   \nonumber \\
&&\{\widetilde{A}^{a,i}(x), \widetilde{\pi}_{j}^b(y) \}|_{\Phi=0}
= \delta^{ab} \delta^{i}_{j}  \delta ({\bf x}-{\bf y})
=\{{A}^{a,i}(x), {\pi}_j^{b}(y) \}_{D},                \nonumber \\
&&\{\widetilde{A}^{a,0}(x), \widetilde{\pi}_{j}^b(y) \}|_{\Phi=0}
=\frac{1}{m^2} g f^{abc} {\pi}^c_j (x)\delta ({\bf x}-{\bf y})
=\{{A}^{a,0}(x), {\pi}_j^{b}(y) \}_{D},                 \nonumber \\
&&\{\widetilde{A}^{a,\mu}(x), \widetilde{\pi}_{0}^b(y) \}|_{\Phi=0}
=0=\{{A}^{a,\mu}(x), {\pi}_0^{b}(y) \}_{D}.
\end{eqnarray}
In this case the function $K$ in Eqs.(38) and (39) corresponds to the 
Dirac brackets $\{A, B \}_{D}$, and hence $\tilde{K}$ corresponding to 
$\{\tilde{A}, \tilde{B} \}$ becomes
\begin{eqnarray}
\{\tilde{A}, \tilde{B} \} = \{ A, B \}_{D} |_{A \rightarrow \tilde{A}, B \rightarrow \tilde{B}}
\end{eqnarray}
which proving our asserted two properties.

Now, since in the Hamiltonian formalism the first class constraint system
without the CS like term [13] indicates the presence of a local symmetry,
this  completes  the  operatorial conversion of the original
second class system with the Hamiltonian $H_T$
and the constraints $\Theta^a_{\alpha}$
into first class one with the Hamiltonian $\widetilde{ H_T}$
and the constraints $\widetilde{\Theta}^a_{\alpha}$.
From Eqs. (18) and (41),
one can easily see that the original second class constraint system
is converted into the effectively first class one if one introduces two fields, $\Phi^1_a, \Phi^2_b$,
which are conjugated with each other in the extended phase space.
Note that for the Proca case
the origin of the second class constraint
is due to the explicit gauge symmetry breaking term in the action (1).

\vspace{1cm}
\begin{center}
{\large \bf IV. Corresponding first class Lagrangian: classical analysis}
\end{center}

Now, let us consider the Lagrangian (first class Lagrangian) corresponding 
to the first class Hamiltonian $\tilde{H}_T$ (or $\tilde{H}_T$). It is 
conjectured that our first class Lagrangian must be related with the 
generalized St\"uckelberg Lagrangian [19] which is gauge invariant by 
construction by noting the relation of the first class constraints and 
the gauge invariance (generated by the first class constraint) according 
to the Dirac conjecture [1]. To confirm this conjecture which being highly 
nontrivial in our case due to the infinitely many terms involved with the 
Hamiltonian and Lagrangian is the object of the section.

To do this, we essentially must perform the inverse Legendre transformation 
by using the Hamilton's equations of motion as will be done in this section. 
But, before this traditional approach, we first consider the approach which showing the 
Lagrangian form more directly by considering the BFT Lagrangian formulation newly.

\begin{center}
{\bf A. BFT Lagrangian formulation}
\end{center}

In the previous sections, we have only concerned about the Hamiltonian
formulation of the BFT method by considering the first class Hamiltonian
$\tilde{H}_c$ or $\tilde{H}_{T}$ which can be easily calculated by replacing ${\cal F}$ by
$\tilde{{\cal F}}$ according to the property (38). However we note that it is not
clear this direct procedure of obtaining the first class quantities
$\widetilde{K}[{\cal F}; \Phi]$ can be valid for the Lagrangian since in this case
the time derivatives of the BFT physical variables are involved which are not
clear to be the BFT physical variables also contrast to the spatial derivatives of the
BFT physical variables. However, interesting enough the time derivatives of the BFT
variables remain BFT physical variables such that the procedure can be 
also applied to the Lagrangian also.

To see this, let us consider the Poisson bracket between time
derivatives of the BFT physical variables with the effectively first class constraints
as follows
\begin{eqnarray}
\{\partial^{0} \tilde{{\cal F}}(x), \tilde{\Theta}_{\beta}(y) \} &=&
  \left\{ \{\tilde{{\cal F}}(x), \tilde{H}_{c} \} + \int d^3 z \tilde{V}(x,z) \tilde{\Theta}_{1}(z), \tilde{\Theta}_{\beta}(y) \right\} ~~~~~~~~~(\beta =1,2)   \nonumber \\
 &=& \left\{ \{\tilde{{\cal F}}(x), \tilde{H}_{c} \} , \tilde{\Theta}_{\beta}(y)\right\}   \nonumber \\
 &=&-\left\{ \{ \tilde{H}_{c},\tilde{\Theta}_{\beta}(y)\}, \tilde{{\cal F}}(x) \right \}
   -\left\{ \{ \tilde{\Theta}_{\beta}(y),\tilde{{\cal F}}(x)\},\tilde{H}_{c} \right \} \nonumber \\
 &=&0,
\end{eqnarray}
where we have used the explicit form, i.e., strong form of the time derivatives as
\begin{eqnarray}
\partial_0 \tilde {\cal F}(x) &=& \{ \tilde{{\cal F}}, \tilde{H}_{T} \} \nonumber \\
 &=&\{\tilde {\cal F}(x), \tilde
H_{c} \} + \int d^3 z \tilde V(x,z) \tilde{\Theta_1}(z), \\
\tilde{V}(x,z) &=&\{\tilde{\cal F}(x), \tilde{u}(z) \} \nonumber
\end{eqnarray}
in the first line and using in the second line the fact that 
$\tilde V$ should be function 
of only the BFT variables since the constituents of $\tilde{u}$ and 
$\tilde{\cal F}$ should be all BFT physical variables and no explicit 
${\cal F}$ dependence in the Poisson algebra of the fundamental BFT 
physical variables as (52). Furthermore we have also
used the Jacobi identity in the third line of Eq. (55). However, this
proof can be also performed by using the strong form of the time 
independence of the constraints (18). We note that this result that the 
time derivatives of the BFT physical variables become also the BFT 
physical variables is universal for any theory due to essentially the 
general validity of the formula (55).

Now, by using this powerful property we can directly obtain the first class 
Lagrangian corresponding to the first class Hamiltonian without using the 
Hamilton's equations of motion. To this end, we first note that due to the 
property (55), the BFT variables $\tilde{\pi}_i$ corresponding to 
$\pi_i =F_{i0}=\partial_i A_0 -\partial_0 A_i -ig [ A_i, A_0]$ is
$\pi ({\cal F})|_{{\cal F} \rightarrow \tilde{\cal F}}$, i.e.,
\begin{eqnarray}
\widetilde{\pi}_i^a [{\cal F}; \Phi] \equiv \widetilde{F}_{i0}^a [{\cal F}; \Phi]= \partial_{i} \widetilde{A}_0^a-\partial_{0} \widetilde{A}_i^a +g f^{abc}\widetilde{A}_i^b \widetilde{A}_0^c = F_{i0}^a [\widetilde{{\cal F}}]. 
\end{eqnarray}

Then, by noting the forms of (31) and (32) which looks like tiled field as 
gauge transformed one with gauge transformation matrix $W$, this form (57) is 
consistent with those forms (31) and (32) only when $\tilde{A}_0^a$ is 
also expressed like as the gauge transformed one as follows
\begin{eqnarray}
\widetilde{A}_{0}^a[{\cal F}; \Phi^{\alpha}]= -\frac{i}{g}\left( \partial_0 W \cdot W^{-1} \right)^a + \left( W A_{0} W^{-1} \right)^a.
\end{eqnarray}
Of course, if the equality (58) is satisfied up to the constraint terms, 
this is also the case 
for (58). Moreover, we note that by comparing the form of (26), (58) can be 
explicitly checked by considering the time derivative of $W$ through the Hamilton's 
equation of motion. Hence together with the forms of (31)-(33), 
we obtain the covariant expressions
\begin{eqnarray}
\widetilde{A}_{\mu}^a[{\cal F}; \Phi]&=& -\frac{i}{g}\left( \partial_{\mu} W \cdot W^{-1} \right)^a + \left( W A_{\mu} W^{-1} \right)^a, \\
\widetilde{F}_{\mu \nu}^a[{\cal F}; \Phi]&=&  \left( W F_{\mu \nu}[{\cal F}] W^{-1} \right)^a                           
= F_{\mu \nu}^a [ \widetilde{{\cal F}}]
\end{eqnarray}
for the four gauge fields and the field strength tensor, respectively. In this form, 
the gauge invariance of the BFT physical variables are manifest.

Now, using this interesting result, we can consider the first
class Lagrangian which is strongly involutive by just replacing ${\cal F}$ by $\tilde{{\cal F}}$
as follows
\begin{eqnarray}
\tilde{\cal L} & \equiv & {\cal L} [\tilde{{\cal F}}] \nonumber \\
    &=& Tr \left[-\frac{1}{2} (W {F}_{\mu\nu}[{{\cal F}}] W^{-1})
    (W {F}^{\mu\nu}[{{\cal F}}]W^{-1} ) \right. \nonumber \\
    &&\left. + m^2 \left(-\frac{i}{g}
    (\partial_{\mu} W) W^{-1} +W A_{\mu} W^{-1} \right)
    \left(-\frac{i}{g} (\partial^{\mu} W) W^{-1}
    +W A^{\mu} W^{-1} \right) \right]   \nonumber \\
    &=& Tr \left[-\frac{1}{2} {F}_{\mu\nu}[{{\cal F}}]
    {F}^{\mu\nu}[{{\cal F}}] + m^2
    \left(A_{\mu}-\frac{i}{g}W^{-1} (\partial_{\mu} W)   \right)
    \left(A^{\mu}-\frac{i}{g}W^{-1} (\partial^{\mu} W)   \right)
   \right],
\end{eqnarray}
where we have used the properties (59) and (60) and the normalization
$Tr(T^a T^b)=\frac{1}{2} \delta^{ab}$ for group generator $T^a$. This provides
the desired Lagrangian corresponding to the first class Hamiltonian
$\tilde{H}_{c}$ or $\tilde{H}_{T}$.

On the other hand, the first class Lagrangian $\tilde{\cal L}$ can be also
obtained by the inverse Legendre transformation of $\tilde{H}_{c}$ or
$\tilde{H}_{T}$ as follows
\begin{eqnarray}
\tilde{\cal L} &=& \tilde{\pi}^{a}_{i} \dot{\tilde{A}}^{a,i}
-\tilde{\cal H}_{c} \nonumber \\
&=& \tilde{\pi}_{\mu}^{a} \dot{\tilde{A}}^{a,\mu} -\tilde{\cal H}_{T}
\end{eqnarray}
with the Hamiltonian density $\widetilde {\cal H}_{c}$ and $\widetilde 
{\cal H}_{T}$ 
\begin{eqnarray}
\tilde{\cal H}_T=\tilde{\cal H}_c -\partial_i \tilde{A}^{a~i} \tilde{\Theta}^{a~i}_1
\end{eqnarray}
corresponding to the Hamiltonian $\widetilde H_{c}$ and 
$\widetilde H_{T}$ of Eqs. (3) and (4), respectively.
Here we have used (60) and the property
\begin{eqnarray}
 \partial_{\mu} \tilde{A}^{a,\mu} =0,
\end{eqnarray}
which can be obtained by replacing ${\cal F}$ by $\tilde{{\cal F}}$ in the equation
\begin{eqnarray}
 \partial_{0} A^{a,0} (x)= \{ A^{a,0}(x), H_{T} \}= -\partial_{i} A^{a,i} (x).
\end{eqnarray}
Here, it is important to note that the inverse Legendre transformation is 
involved only the BFT variables because the first class Hamiltonian $\tilde{H}_T$ 
or $\tilde{H}_c$ are expressed solely by the BFT variables.  
Furthermore, we note that the first class Lagrangian $\tilde{\cal L}$ has been
reduced to the so-called the generalized St\"uckelberg Lagrangian [19]
\begin{eqnarray}
{\cal L}_{St\ddot{u}ckelberg}  = Tr \left[-\frac{1}{2} {F}_{\mu\nu}
    {F}^{\mu\nu} + m^2
    \left(A_{\mu}-\frac{i}{g}W^{-1} (\partial_{\mu} W)   \right)
    \left(A^{\mu}-\frac{i}{g}W^{-1} (\partial^{\mu} W)   \right)
   \right]
\end{eqnarray}
in the last line of Eq.(62) after in the trace operation with the identification of the
field $\Phi^2$ with the St\"uckelberg's scalar such that the St\"uckelberg's
formulation can be understood as an
Lagrangian formulation of the BFT method. However, it seems appropriate to 
comment
that, contrast to Abelian model, the simple replacement of
\begin{eqnarray}
A_{\mu} \rightarrow A_{\mu} -\frac{i}{g} W^{-1} \partial_{\mu} W
\end{eqnarray}
in the Lagrangian ${\cal L}[{\cal F}]$ of Eq. (63) inspired by the mass term
of (66) does
not reproduce the generalized St\"uckelberg Lagrangian since in that case the kinetic
term $Tr \frac{1}{2}(F_{\mu \nu} F^{\mu \nu})$ is not invariant under this
replacement (67) but produces several additional terms. In this sense our
understanding the St\"uckelberg Lagrangian as the original Lagrangian form in
the BFT variables space is more simple and hence can be considered as the more
fundamental formulation.

\begin{center}
{\bf B. Lagrangian from the Hamilton's equation of motion}
\end{center}

We now derive the same first class Lagrangian ${\tilde{\cal L}}$ from the traditional approach which is using
the Hamilton's equations of motion [21].
However, as we shall see, some slight modification in the BFT Hamiltonian 
$\tilde{H}_T$ (or $\tilde{H}_c$) should be needed in order to realize 
the result of (58).

In order to understand the idea of this modification, we first consider the relatively simple 
case of Abelian theory without modification which is described by the first class 
Hamiltonian
\begin{equation}
\tilde{H}_T = \int d^3x \left[
                \frac{1}{2} \tilde{\pi_i}^2 + \frac{1}{4} \tilde{F_{ij}}^2
              + \frac{1}{2} m^2 (\tilde{A^0}^2 + \tilde{A^i}^2 )
              - \tilde{A_0} \tilde{\Theta_2}-\partial_i \tilde{A^i} \tilde{\Theta}_1
              \right].
\end{equation}
Then, the Hamilton's equation of motion for $\tilde{A}_i$ is resulted to be
\begin{eqnarray}
\dot{\tilde{A}_i} \equiv \{\tilde{A}_i, \tilde{H}_T \}=-\tilde{\pi}_i +\partial_i \tilde{A}_0 -\frac{1}{m^2} \partial_i \tilde{\Theta}_2
\end{eqnarray}
such that the equality corresponding to the Abelian version of (57) is satisfied only weakly, i.e.,
\begin{eqnarray}
\tilde{\pi}_i =\partial_i \tilde{A}_0 -\partial_0 \tilde{A}_i-\frac{1}{m^2} \partial_i \tilde{\Theta}_2.
\end{eqnarray}
Here, the relation corresponding to (58) is not satisfied strongly but only 
up to non-tilled Gauss law constraint $\Theta_2$
\begin{eqnarray}
\tilde{A}_0 =A_0 +\frac{1}{m} \partial_0 \Phi^2 -\frac{1}{m^2}\Theta_2.
\end{eqnarray}
Now, in order to realize (58) in a strong form we must consider the modified 
Hamiltonian
\begin{eqnarray}
\tilde{H}^*_T =\tilde{H}_T +\int d^3 x \frac{\Phi^1}{m} \tilde{\Theta}_2
\end{eqnarray}
which has been originally introduced in other context by us recently [16]. With this Hamiltonian as 
the true time translation generator, we obtain
\begin{eqnarray}
\dot{\tilde{A}_i} \equiv \{\tilde{A}_i, \tilde{H}^*_T \}=-\tilde{\pi}_i +\partial_i \tilde{A}_0 
\end{eqnarray}
such that we obtain the strong equality
\begin{eqnarray}
\tilde{\pi}_i =\partial_i \tilde{A}_0 -\partial_0 \tilde{A}_i
\end{eqnarray}
and hence the strong equality for the gauge field $\tilde{A}_0$
\begin{eqnarray}
\tilde{A}_0 =A_0 +\frac{1}{m} \partial_0 \Phi^2 
\end{eqnarray}
or by using the the Abelian result of $\tilde{A}_i$ of (31) the the four gauge fields
are found to be
\begin{eqnarray}
\tilde{A}_{\mu} =A_{\mu} +\frac{1}{m} \partial_{\mu} \Phi^2.
\end{eqnarray}
Now, with this strong equality we can easily obtain the
St\"uckelberg action by considering the inverse Legendre transformation 
(62). 
Interestingly enough, we note that the time evolution of the constraints 
generated by $\tilde{H}^*_T $ is isomorphic to the original ones as follows
\begin{eqnarray}
&&\dot{\tilde{\Theta}}_1=\{ \tilde{\Theta}_1, \tilde{H}^*_T \} =\tilde{\Theta}_2, \nonumber \\
&&\dot{\tilde{\Theta}}_2=\{ \tilde{\Theta}_2, \tilde{H}^*_T \} =0.
\end{eqnarray}
However, we note that this modified Hamiltonian does not change the 
physically observable because the modification term affect the original
Hamilton's equations of motion only by the terms proportional to the first class
constraints also as follows
\begin{eqnarray}
\{\tilde{\cal F}(x), \tilde{H}^*_T \} =\{\tilde{\cal F}(x), \tilde{H}_T \}
+\int d^3 y \{ \tilde{\cal F}(x), \frac{\Phi^1}{m}(y) \} \tilde{\Theta}_2 (y).
\end{eqnarray}
This is the idea of the Hamiltonian modification.

Now, let us return to the non-Abelian case which is described by the Hamiltonian
(44) before modification. Similar to the Abelian case, the Hamilton's equations 
of motion are resulted to be
\begin{eqnarray}
\dot{\tilde{A}_i^a} \equiv \{\tilde{A}_i^a, \tilde{H}_T \}=-\tilde{\pi}_i^a +\tilde{D}_i \tilde{A}_0^a -\frac{1}{m^2} (\tilde{D}_i \tilde{\Theta}_2)^a
\end{eqnarray}
such that the equality corresponding to (57) is satisfied only weakly, i.e.,
\begin{eqnarray}
\tilde{\pi}_i^a =\partial_i \tilde{A}_0^a -\partial_0 \tilde{A}_i^a+g f^{abc}\tilde{A}^b_i \tilde{A}_0^c-\frac{1}{m^2} (\tilde{D}_i \tilde{\Theta}_2)^a.
\end{eqnarray}
Hence, in order to realize (58) in a strong form we must consider the modified 
Hamiltonian
\begin{eqnarray}
\tilde{H}^*_T &=&\tilde{H}_T  \nonumber \\
      && +\int d^3 x \left\{
\frac{1}{m} \Phi^{1}_a -\sum_{n=1}^{\infty} \frac{n}{(n+1)!} \frac{1}{m^2}
  \left(\frac{g}{m} \right)^{n} [\Phi^{2} \times ( \Phi^{2} \times ( \cdots ( \Phi^{2} \times \Omega )\cdots)) ]_{\mbox{n-fold}}^{a}
 \right\} \tilde{\Theta}^a_2 \nonumber \\
&=& \tilde{H}_T + \int d^3 x (\tilde{A}^a_0 -A^a_0) \tilde{\Theta}^a_2
\end{eqnarray}
which produces the strong equality (57) and hence our desired form (58) 
or (59).
The Abelian case (72) is easily reduced form this general non-Abelian formula.
Then, in this case we can also easily obtain the generalized St\"uckelberg
Lagrangian as in the previous subsection {\bf A}. Furthermore, with this 
modification, we can also have the same evolution of the constraint as the 
original ones as
\begin{eqnarray}
&&\dot{\tilde{\Theta}}_1^a=\{ \tilde{\Theta}_1^a, \tilde{H}^*_T \} =\tilde{\Theta}_2^a, \nonumber \\
&&\dot{\tilde{\Theta}}_2^a=\{ \tilde{\Theta}_2^a, \tilde{H}^*_T \} =0.
\end{eqnarray}

\begin{center}
{\large \bf V. Conclusion}
\end{center}

In conclusion, we have applied the newly improved BFT Hamiltonian and
Lagrangian formalism, which provide the more simple and transparent insight
to the usual BFT method, to the non-Abelian Proca model as an non-trivial
application of our new method and completely solved the model by the
ingenious choice of the antisymmetric $X_{\alpha \beta}, \omega^{\alpha \beta}$
as we have suggested recently [16].

Firstly, by applying the usual BFT formalism [7] to the non-Abelian
Proca model, we have converted the weakly second class constraint system
into a strongly first class one by introducing new auxiliary fields. As a result we 
have  found that the effectively first class constraints needs the infinite terms involving
the auxiliary fields which being regular power series form in our ingenious choice.

Moreover, according to our new method, the first class Hamiltonian, which also
needs infinite correction terms, is obtained simply by replacing the original
variables in the original Hamiltonian with the BFT physical variables. The
effectively first class constraints can be also understood similarly. Furthermore, we have shown
that the infinite terms in the BFT physical variables and hence first
class constraints and the first class Hamiltonian can be expressed by
the exponential involved form. On the other hand, we  have also shown that in our
model the Poisson brackets of the BFT physical variables fields in the
extended phase space are the same as the Dirac brackets of the phase space
variables in the original second class constraint system  only replacing
the original phase variables by the BFT physical variables. We note that
this result is conformity with the general result of BFT [7] and hence support
the consistency of our result.

We have also newly developed the BFT Lagrangian formulation, which has not been
unclear so far [16, 17], to obtain the first
class constraints which complements and provides much more transparent insight
to the Hamiltonian formulation. Based on this BFT Lagrangian formulation we
directly obtain the St\"uckelberg Lagrangian as the classical Lagrangian 
corresponding to the first class Hamiltonian which is most non-trivial conjecture 
related to Dirac's conjecture [1] in our infinitely many terms involved model. This is also confirmed by
considering the Hamilton's equations of motion with some modification term in 
Hamiltonian proportional to  the first class constraints $\tilde{\Theta}_2^a$.
This result disproves the recent argument on the in-equivalence of
St\"uckelberg formalism and BFT formalism.

As for the other applications of our new method, we have found that our method
is also powerful in the system with both the second and first class
constraints like as the non-Abelian Chern-Simons gauge theory [24, 25].
Moreover, the investigation on the corresponding dual theory [21, 26] which would
be the non-Abelian Kalb-Ramond field interacting with the Yang-Mills 
field  would be interesting. Furthermore, we note that recently our 
improved BFT method is
found to be also powerful in the path integral equivalence of the 
Maxwell-Chern-Simon theory with the Abelian self-dual model [27] and the BFT
formulation of the non-Abelian self-dual model [28].

Finally, we would like to comment on the interesting work recently done by
Banerjee and  Barcelos-Neto [29]. After finishing our work, we have found that
their work is similar to our work. However, we  have found that their work is
still incomplete compared
to ours in that firstly they had not found our compact forms (31) and (32)
and hence could not arrange explicitly the complicate first class
Hamiltonian (41). Furthermore, they did not completely determine $A_0^a$
contrast to our complete determination in Eq. (26). Nevertheless, it is interesting to
note that they have succeed in proving the equivalence of BFT Hamiltonian
formalism and the generalized St\"uckelberg's Lagrangian formalism by
identifying the constraints of both formalism order by order. This can be
considered as another independent proof of our results of section IV, where
we have prove the result more compactly. Moreover, we have directly read off
the Dirac bracket from the
Poisson brackets of the BFT physical variables.

\vspace{1cm}
\begin{center}
{\large \bf ACKNOWLEDGMENTS}
\end{center}
We would like to thank W. T. Kim, Y.-W. Kim, B. H. Lee, K.D. Rothe for
helpful discussions.
The present work was supported by the Basic Science Research Institute
program, Ministry of Education, 1996, Project No. BSRI-96-2414.

\newpage
\vspace{1cm}
\begin{center}
{\large \bf REFERENCES}
\end{center}

\begin{description}

\item{[1]} P. A. M. Dirac, `` Lectures on quantum mechanics ''
         ( Belfer graduate School,
            Yeshiba University Press, New York, 1964.
\item{[2]} E. S. Fradkin and G. A. Vilkovisky,
            {\it Phys. Lett. B} {\bf 55} (1975), 224.
\item{[3]} M. Henneaux, {\it Phys. Rep. C} {\bf 126} (1985), 1.
\item{[4]} C. Becci, A. Rouet and R. Stora,
           {\it Ann. Phys. (N.Y.) } {\bf 98} (1976), 287;
           I. V. Tyutin, Lebedev Preprint 39 (1975).
\item{[5]} T. Kugo and I. Ojima,
           {\it Prog. Theor. Phys. Suppl.} {\bf 66} (1979), 1.
\item{[6]} I. A. Batalin and E. S. Fradkin,
           {\it Nucl. Phys. B} {\bf 279} (1987), 514;
           {\it Phys. Lett. B} {\bf 180} (1986), 157.
\item{[7]} I. A. Batalin and I. V. Tyutin,
           {\it Int. J. Mod. Phys. A }{\bf 6} (1991), 3255; E. S. Fradkin,
           `` Lecture of the Dirac Medal of ICTP 1988 '',  Miramare-Trieste
           , 1988.
\item{[8]} T. Fujiwara, Y. Igarashi and J. Kubo,
            {\it Nucl. Phys. B} {\bf 341} (1990), 695;
            {\it Phys. Lett. B} {\bf 251} (1990), 427; J. Feinberg and
            M. Moshe, {\it Ann. Phys.} {\bf 206} (1991), 272.
\item{[9]} Y.-W. Kim, S.-K. Kim, W. T. Kim, Y.-J. Park,
           K.Y. Kim, and Y. Kim, {\it Phys. Rev. D} {\bf 46} (1992), 4574.
\item{[10]} R. Banerjee, H. J. Rothe and K. D. Rothe,
           {\it Phys. Rev. D} {\bf 49} (1994), 5438.
\item{[11]} L. D. Faddeev and S. L. Shatashivili,
           {\it Phys. Lett. B} {\bf 167} (1986), 225;
           O. Babelon, F. A. Shaposnik and C. M. Vialett,
           {\it Phys. Lett. B} {\bf 177} (1986), 385;
           K. Harada and I. Tsutsui, { \it Phys. Lett. B} {\bf 183} (1987), 311.
\item{[12]} J. Wess and B. Zumino, {\it Phys. Lett. B} {\bf 37} (1971), 95.
\item{[13]} R. Banerjee, {\it Phys. Rev. D} {\bf 48} (1993), R5467;
           Y.-W. Kim, Y.-J Park, K. Y. Kim and Y. Kim, {\it Phys. Rev. D}
           {\bf 51} (1995), 2943;
           E.-B. Park, Y.-W. Kim, Y.-J Park, Y. Kim, and W. T. Kim,
           {\it Mod. Phys. Lett. A} {\bf 10} (1995), 1119.
\item{[14]} N. Banerjee, S. Ghosh and R. Banerjee,
           {\it Nucl. Phys. B} {\bf 417} (1994), 257;
           {\it Phys. Rev. D} {\bf 49} (1994), 1996; R. Amorim and J. Barcelos. Neto,
           {\it Phys. Lett. B} {\bf 333} (1994), 413.
\item{[15]} N. Banerjee, R. Banerjee and S. Ghosh, {\it Ann. Phys.}
           {\bf 241} (1995), 237.
\item{[16]} Y.-W. Kim, M.-I. Park, Y.-J. Park, and S. J. Yoon, {\it Int. J. Mod.
           Phys. A} (to be published) (hep-th/9702002); M.-I. Park and Y.-J Park, {\it New Physics} ( to be published).
\item{[17]} W. T. Kim, Y.-W. Kim, M.-I. Park, Y.-J. Park, and S. J. Yoon,
          {\it J. Phys. G}23 (1997), 325.
\item{[18]} P. Senjanovic, {\it Ann. Phys. (N.Y.)} {\bf 100} (1976), 227;
          {\bf 209} (1991), 248(E).
\item{[19]} T. Kunimasa and T. Goto, {\it Prog. Theor. Phys.} {\bf 37} (1967), 452;
 A. S. Slavnov, {\it Theor. Math. Phys.} {\bf 10} (1972), 305;
 K. Shizuya, {\it Nucl. Phys. B} {\bf 94} (1975), 260; C. Grosse-Knetter, {\it Phys. Rev. D}
 {\bf 48} (1993), 2854.
\item{[20]} R. Amorim and A. Das, {\it Mod. Phys. Lett. A} {\bf 9} (1994), 3453;
              R. Amorim, {\it Z. Phys. C} {\bf 67} (1995), 695.
\item{[21]} N. Banerjee and R. Banerjee, {\it Mod. Phys. Lett. A} {\bf 11} (1996),
              1919.
\item{[22]} T. Maskawa and H. Nakajima, { \it Prog. Theor. Phys.} {\bf 56} (1976), 1295;
     S. Weinberg, { `` The Quantum Theory of Fields '' }, Cambridge University Press, New York, 1995.
\item{[23]} L. D. Faddeev and V. N. Popov, {\it Phys. Lett. B} {\bf 25} (1967), 29.
\item{[24]} W. T. Kim and Y.-J. Park, {\it Phys. Lett. B} {\bf 336} (1994), 376.
\item{[25]} M.-I. Park and Y.-J. Park, `` Quantization of the Chern-Simons 
theories based on the improved BFT formalism ", Sogang University Report No. 
SOGANG-HEP 213/97, 1997 (unpublished).
\item{[26]} R. Banerjee, {\it Nucl. Phys. B} {\bf 465} (1996), 157.
\item{[27]} R. J. Banerjee, H.J. Rothe and K. D. Rothe, {\it Phys. Rev. D} {\bf 55} (1997), 6339.
\item{[28]} Y.-W. Kim and K. D. Rothe, " BFT Hamiltonian embeeding of Non-Abelian
self-dual model", Heidelberg University Report No. HD-THEP-97-25 (hep-th/9706018),
 1997 (unpublished). 
\item{[29]} R. Banerjee and J. Barcelos-Neto, `` Hamiltonian embedding of the
            massive Yang-Mills theory and the generalized St\"uckelberg
            formalism'' (hep-th/9701080), 1997 (unpublished).
\end{description}
\end{document}